\documentclass[11pt]{article}
\usepackage{graphicx}
\usepackage{multirow}
\usepackage{array}
\usepackage{float}
\usepackage{soul}
\usepackage[
    left=2cm,
    right=2cm,
    top=2.5cm,
    bottom=2.5cm
]{geometry}
\usepackage[numbers,sort&compress]{natbib}
\usepackage{microtype}
\usepackage{hyperref}
\usepackage{booktabs}
\usepackage{amssymb}
\usepackage{pifont}
\usepackage{xcolor}
\usepackage{marvosym}

\newcommand{\cmark}{\ding{51}} 
\newcommand{\xmark}{\ding{55}}

\begin{document}

\title{The BEATS Dataset: Towards Understanding Music-Physiology Dynamics in Real-World Scenarios}

\author{
Benjamin Gutierrez Serafin$^{1}$\textsuperscript{\Letter},
Tanaya Guha$^{1}$,
Fahim Kawsar$^{1,2}$ \\
\\
$^{1}$University of Glasgow, United Kingdom \\
$^{2}$OmniBuds Ltd, United Kingdom \\
\textsuperscript{\Letter} Corresponding author:
\href{mailto:b.gutierrez-serafin.1@research.gla.ac.uk}{\texttt{b.gutierrez-serafin.1@research.gla.ac.uk}}
}

\date{}

\maketitle

\begin{abstract}
We introduce the BEATS dataset, a robust compilation of psycho-physiological data collected in the wild from 30 participants while they listened to curated playlists intended to promote well-being. An acronym for Biomarkers, Experience and Affect from Therapeutic Soundscapes, the BEATS dataset aims to support research on wellbeing, music psychology, affective computing and signal processing. More specifically, this dataset includes: (1) sensor data collected during music listening sessions using a research-graded wearable device in free-movement environments; (2) music listening activity records, stimulus metadata, acoustic features, and self-reports associated with the listening sessions (3) individual characteristics, such as demographics, music preference, musical skills, and personality traits; and (4) self-reported psychological measures before and after their experimental participation spanning for a number of weeks. By situating data collection outside of lab settings, the BEATS dataset offers the opportunity to study real-world music-listening scenarios, allowing for a richer exploration of the physiological, subjective and therapeutic effects of music under real-world conditions.
\end{abstract}

\section{Background and Summary}
Music has been shown to have significant therapeutic value across a wide range of health and well-being applications, including mood and stress regulation, pain reduction, memory support, and even improving neurological and mental health conditions such as Parkinson’s disease and eating disorders \cite{chong2024scoping, de2022music, sihvonen2022isn, testa2020systematic, matziorinis2022promise, levitin2025music}. Due to its non-invasive nature, lack of adverse side effects, and relatively low cost, music can often be an attractive alternative to pharmacological interventions \cite{kakar2021music}. However, it remains unclear how musical characteristics (e.g., pitch, tempo, timbre) affect physiological responses associated with these therapeutic benefits \cite{hernandez2020makes, chee2024effects}. Existing studies do not yet offer clear insights, limiting music-based interventions from delivering desirable therapeutic outcomes.

Current evidence links music listening to physiological regulation and beneficial changes in affective states \cite{zaatar2024transformative, liu2026psychological, raglio2026music}. Because emotional experiences modulate the autonomic nervous system activity, they produce measurable changes in bodily responses such as heart rate variability (HRV) and skin conductance \cite{egger2019emotion}. Derived from physiological measurements obtained using photoplethysmography (PPG) and electrodermal activity (EDA), these biomarkers provide objective indicators of emotional valence (pleasantness–unpleasantness) and arousal (low–high activation). Therefore, the established relationship between these biomarkers and emotional states \cite{shu2018review, giannakakis2019review, dzedzickis2020human} offers a framework for investigating how music influences physiology.

At the same time, advances in music streaming services and wearable sensing technologies have opened new avenues to collect physiological and music listening data at scale in real-world settings and over extended periods \cite{greenberg2017music}. This paradigm allows the investigation of music–physiology relationships under real-world conditions while simultaneously enabling analysis of intentional music listening behaviours. Multimodal wearable data have been successfully used to infer affective states using traditional machine learning approaches, but mostly within short-term, semi-controlled studies \cite{schmidt2018introducing, iqbal2022stress, rahmani2024emowear, schreiber2025stress}. While field studies pose challenges related to signal quality and ground truth, they enable ecologically valid assessment of emotions and are increasingly feasible with current wearable technologies \cite{saganowski2022emotion}. In parallel, music consumption has become ubiquitous, where listening takes place alongside other activities rather than the primary activity \cite{kassabian2013ubiquitous}. The organisation of music into user-made and platform-generated playlists centred on contexts, moods, and activities illustrates how music listening has become a dispersed practice that supports, regulates, or enriches other main activities.\cite{walsh2024streaming, fuentes2019soundtracking}.

Traditionally, music psychology and clinical trials employing receptive listening approaches rely on diverse and study-specific stimulus selection strategies tailored to the specific aims and hypotheses under investigation. Music may be self-selected by participants either from their personal music collection \cite{kakde2023effect, kiss2021effect} or a predefined pool of stimuli \cite{nguyen2025randomised, wakana2022effect}, curated by experts or chosen in collaboration with participants \cite{mcferran2022receptive}, drawn from predefined catalogues \cite{mir2021relaxing, durgun2023randomized}, or specifically composed or adapted for the study \cite{gonzalez2019more, hernandez2020makes}. However, the lack of standardisation of stimulus sets and selection criteria, which are often not described exhaustively, limits methodological comparability and replicability across investigations \cite{tan2012interplay}. 

Most studies on music-evoked emotions have been conducted in constrained laboratory settings, which can skew physiological responses to music and limit their transferability to real-world listening scenarios \cite{jaimovich2013emotion, linnemann2016stress, linnemann2017assessing}. Moreover, experimental designs often overlook individual differences (e.g., personality and music preferences), or the listener’s intention to engage with music, both of which are known to shape emotional responses \cite{juslin2022emotions}. Existing evidence is also largely limited to short temporal snippets rather than longer term behaviour i.e., listening to long playlist of music across multiple days \cite{butkow2024eartune, saskovets2025effects}. While the above studies show that musical features can predict short-term physiological responses, these effects remain to be validated under real-world conditions. Achieving this, however, requires addressing the challenges of ambulatory physiological sensing, such as movement and body posture introducing motion artifacts \cite{norman2026wearable}, environmental conditions causing EDA to change due to thermoregulation rather than affective states \cite{raugh2019comprehensive}.

Despite the maturity of wearable sensing, datasets designed to study the effect of music on physiology in naturalistic settings remain largely absent. Table 1 summarises all publicly available datasets in this domain so far. We note that they all rely on short excerpts of popular songs, often with vocals or visual content, making it difficult to determine whether observed physiological responses are driven by the acoustic characteristics of the stimuli or by their semantic or visual components \cite{koelstra2011deap, zhang2018pmemo, hu2022detecting, grigorev2024situnes}. Furthermore, some experimental designs did not ensure that all participants were exposed to the same musical stimuli, making fair comparisons of physiological responses across individuals difficult. Beyond these limitations, research remains dominated by laboratory-based studies, with only one week-long field study limited to heart rate measurements instead of raw multimodal physiological signals necessary to derive multiple biomarkers and model emotional valence and arousal. Taken together, these limitations underscore the need for datasets that capture music-induced changes in physiology and via multimodal wearable sensor data under realistic conditions.

\begin{figure}[ht]
  \centering
\includegraphics[width=0.85\linewidth]{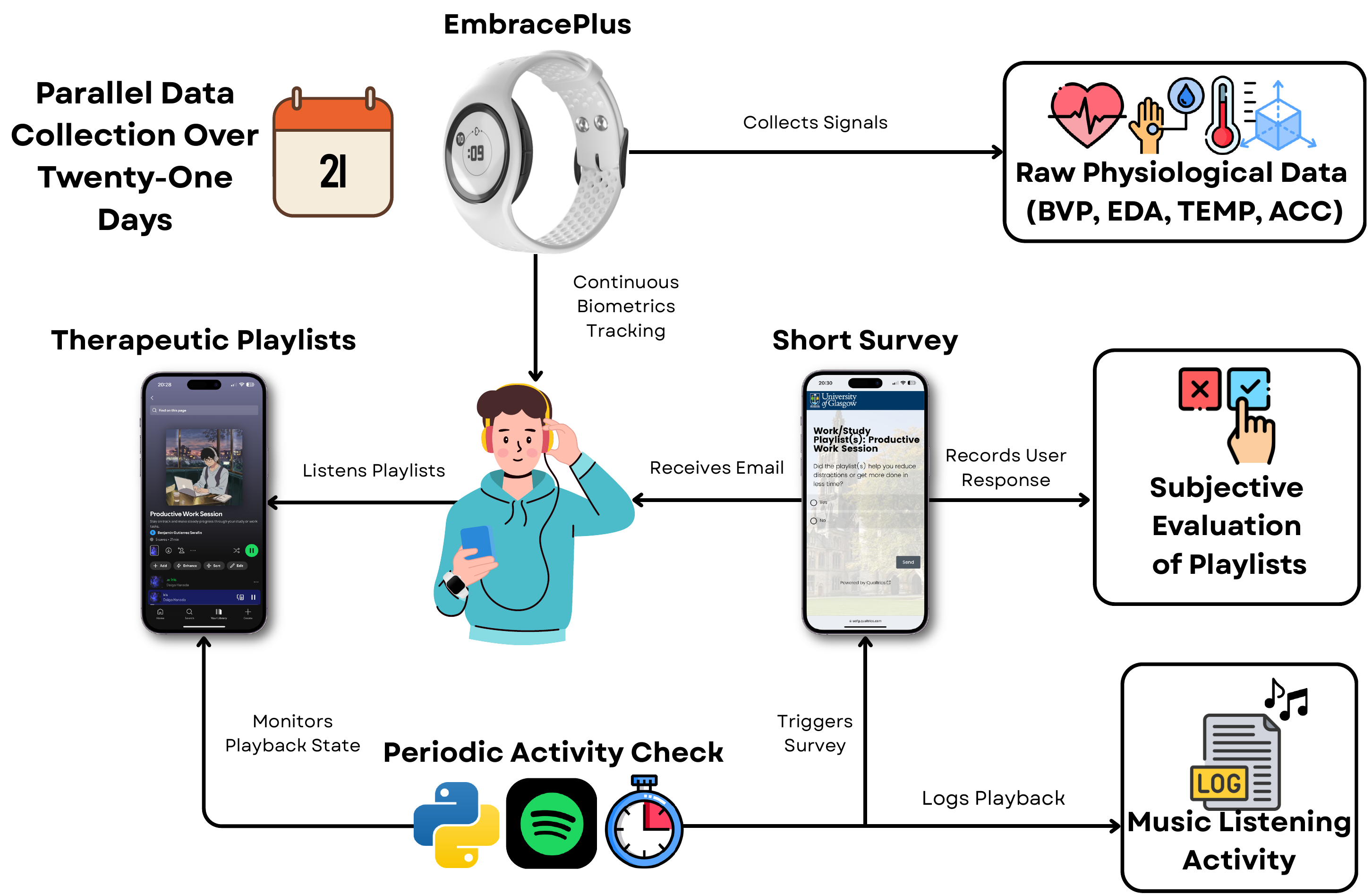}
  \caption{Overview of the in-the-wild data collection procedure, illustrating continuous physiological monitoring, automated logging of music listening activity, and time-triggered email surveys linking subjective evaluations to playback and biometrics over the 21-day study period. }
  \label{fig:procedure}
\end{figure}

\textbf{Summary.} We developed an open-source wearable sensing and music streaming system for in-the-wild collection of music listening and physiological data. Figure 1 provides an overview of the system and data collection workflow. Building on our initial work \cite{gutierrez2026towards}, we introduce the \textbf{BEATS} dataset. It consists of  21-day in-the-wild recordings of multimodal physiological data from 30 participants. The participants were recorded continuously as they listened to music in real-world environments. Over the experimental period, participants continuously wore the \textit{Empatica EmbracePlus}, a research-grade wrist-worn device, throughout their daily routines and voluntarily listened to systematically selected playlists using a dedicated \textit{Spotify} account. The dataset contains raw physiological signals, (PPG, EDA, skin temperature, triaxial acceleration), music listening history, pre-computed acoustic features of the stimuli, subjective assessments of listening experiences (via self-reports), demographic information, individual difference measures (e.g., personality, music preference, and musical skills), and pre/post measures of mood, perceived stress, and depressive symptoms. With more than 10,000 hours of physiological recordings from 30 participants (15 female; aged 18–49), of which 262 hours overlap with music listening, BEATS is the first and the largest datasets of its kind. We expect BEATS to serve as a valuable resource for the community to advance our understanding of music-physiology dynamics and for developing effective personalised music-based interventions.

\begin{table*}[t]
\centering
\footnotesize
\setlength{\tabcolsep}{2pt}
\label{tab:dataset_comparison}
\caption{Summary of available music datasets with temporally aligned physiological data. \dag Samples denote valid music-listening instances that overlap with physiological signals. \ddag Lstn. Ctrl. indicates whether participants controlled the timing, stimulus, and onset of music listening, rather than exposure being experimenter- or system-delivered. \S Indiv. Diff. indicates whether participant characteristics (e.g., personality, familiarity, and music preferences) were collected.}
\vspace{2mm}
\begin{tabular}{
>{\raggedright\arraybackslash}m{1.5cm}
>{\raggedright\arraybackslash}m{2cm}
>{\centering\arraybackslash}m{2cm}
>{\centering\arraybackslash}m{1.3cm}
>{\centering\arraybackslash}m{1.5cm}
>{\raggedright\arraybackslash}m{2.3cm}
>{\centering\arraybackslash}m{1.3cm}
>{\raggedright\arraybackslash}m{1.5cm}
>{\centering\arraybackslash}m{1.5cm}
>{\centering\arraybackslash}m{1cm}
}
\toprule
\textbf{Dataset} & \textbf{Stimuli} & \textbf{Participants} & \textbf{\dag Samples} & \textbf{\ddag Lstn. Ctrl.} & \textbf{Data Collected} & \textbf{\S Indiv. Diff.} & \textbf{Device} & \textbf{Setting} & \textbf{Available} \\
\midrule
DEAP \cite{koelstra2011deap} & 40 music video excerpts (60s) & 32 & 1,280 & \xmark & EEG, GSR, EMG, BVP, face video, self-reports & \xmark & Biosemi ActiveTwo & Lab & \cmark \\
PMEmo \cite{zhang2018pmemo} & 794 pop chorus excerpts (11-88s) & 457 & 7,940 & \xmark & EDA, static/dynamic valence-arousal ratings & \xmark & Biopac MP150 & Lab & \cmark\\
Emotion in Motion (EiM) \cite{jaimovich2013emotion,huang2015exploring, huang2017exploratory, bortz2019cross}  & 53--81 song excerpts ($\sim$90\,s) & $\geq$20,000 & 6,509 & \xmark & EDA, HR, self-reports & \cmark & POX-OEM M15HP and BioEmo & Field & \xmark \\
HKU956 \cite{hu2022detecting} & 956 full-length CC-licensed songs ($\sim$50\,s avg.) & 30 & 956 & \cmark & BVP, EDA, skin temperature, HR, valence/arousal ratings & \cmark & Empatica E4 & Lab & \cmark \\
SiTunes \cite{grigorev2024situnes} & 307 tracks (varied duration) & 30 & 2,006 & \xmark & HR, activity, location, valence/arousal, liking, duration & \xmark & Smart wristband & Field (1 week) & \cmark \\
\textbf{BEATS} & 117 tracks across 3 contexts (varied duration) & 30 & 5,173 & \cmark & BVP, EDA, skin temperature, tri-axial ACC, 1-min biomarkers, binary self-reports & \cmark & Empatica EmbracePlus & Field (21 days) & \cmark \\
\bottomrule
\end{tabular}
\end{table*}

\section{Methods}
To enable physiological sensing beyond short-term laboratory protocols, we collected music listening and physiological data in-the-wild from 30 participants over 21 days using a custom-built open-source framework integrating music streaming and wearable technology. In this section, we describe the study population, the data collection protocol and the collected data. 

\subsection{Participants} Thirty participants were voluntarily recruited in sequential batches, aligned with device availability; data collection commenced in May 2025 and concluding in December 2025. Participants' age ranged from 18 to 49 years (majority between 25 and 34, n = 24) with no auditory issues, with equal gender representation (15 females) and diverse cultural backgrounds. To be eligible, individuals were required to commit to the 21-day study period, willingness to wear the device continuously, and own a compatible mobile device (\href{https://www.empatica.com/embraceplus-compatibility/}
{\nolinkurl{https://www.empatica.com/embraceplus-compatibility/}}) to install the companion smartphone application required for data collection. Prior to enrolment, each participant signed an informed consent form granting permission for their data to be used for research purposes. Participants were compensated with a £50 Amazon voucher upon study completion. Ethical approval was obtained from the University of Glasgow's College of Science and Engineering ethics committee, under reference number 300240032.

\subsection{Stimuli, system and setup} Before participant recruitment, we systematically selected the music resources of the study to constrain variability in listening conditions and ensure repeated exposure to a shared list of stimuli, facilitating comparability across participants. 

Looking at computational approaches to stimulus selection, prior work has employed semi-automated pipelines combining affective tags and online ratings to segment emotionally salient excerpts \cite{koelstra2011deap}, curated popular pop songs sourced from major music charts \cite{zhang2018pmemo}, and participant-driven selection within predefined catalogues in controlled settings \cite{hu2022detecting}. To facilitate the study of music–physiology relationships in real-world conditions, we propose a computationally grounded stimulus selection approach. Guided by the work of Groarke and Hogan \cite{groarke2016enhancing} on identifying functions of music for enhancing well-being, we chose three high-level \emph{musical functions} for stimulus selection. To minimise contextual variability, each was restricted to specific real-world scenarios as follows: 
\begin{itemize}
    \item \textbf{Affect regulation}: Music for mood improvement and stress reduction to promote relaxation.
    \item \textbf{Eudaimonic}: Music for introspection, meditation and personal meaning to facilitate mental clarity.
    \item \textbf{Goal attainment}: Music to increase motivation and persistence in work or study sessions.
\end{itemize}
This functional framing mirrors current streaming services, in which music is curated into context-specific playlists supporting everyday activities. Accordingly, Spotify (\href{https://www.spotify.com/us/}
{\nolinkurl{https://www.spotify.com/us/}}) served as the platform for stimulus selection and delivery. Figure 2 illustrates the pipeline used for stimulus selection. Playlist descriptions were embedded using the \textit{all-mpnet-base-v2} (\href{https://huggingface.co/sentence-transformers/all-mpnet-base-v2}{\nolinkurl{https://huggingface.co/sentence-transformers/all-mpnet-base-v2}}). Candidate playlists $i \in I$ were ranked for each music function $j \in J$ using a weighted relevance score defined as

\begin{equation}
{S}_{i,j} = w_1 P_i + w_2 A_i + w_3 M_{i,j},
\qquad \sum_{k=1}^{3} w_k = 1
\label{eq:wrs}
\end{equation}
where $P_i \in [0,1]$ denotes the min--max normalised logarithm of the number of followers of the $i$-th playlist, $A_i$ the semantic similarity between playlist descriptions and predefined acoustic criteria (e.g., instrumental, no vocals), and $M_{i,j}$ the semantic similarity between playlist descriptions and the keywords describing music function $j$. Semantic similarity scores were computed using cosine similarity between the corresponding text embeddings. We empirically set $w_1=0.3$, $w_2=0.3$, and $w_3=0.4$, as the slightly higher weighting of functional relevance provided better playlist assignments than equal weighting, while retaining contributions from popularity and acoustic suitability. Each playlist was assigned to the music function with the highest weighted relevance score

\begin{equation}
R_{\mathrm{music}}(i)=\arg\max_{j} S_{i,j}
\label{eq:f_music}
\end{equation}

Using this score, playlists were ranked within each music function, and the highest-ranked playlist from each function was selected and divided into four 30-minute playlists per functional category.

% Playlist descriptions were embedded using the \textit{all-mpnet-base-v2} (\href{https://huggingface.co/sentence-transformers/all-mpnet-base-v2}{\nolinkurl{https://huggingface.co/sentence-transformers/all-mpnet-base-v2}}), and cosine similarity between function-specific keywords and playlist embeddings were used to rank candidate playlists, with popularity metrics and predefined musical criteria applied as an additional filter. The top-ranked playlists were then subsampled into four 30-minute playlists per functional category.

\begin{figure}[ht]
  \centering
  \includegraphics[width=0.9\linewidth]{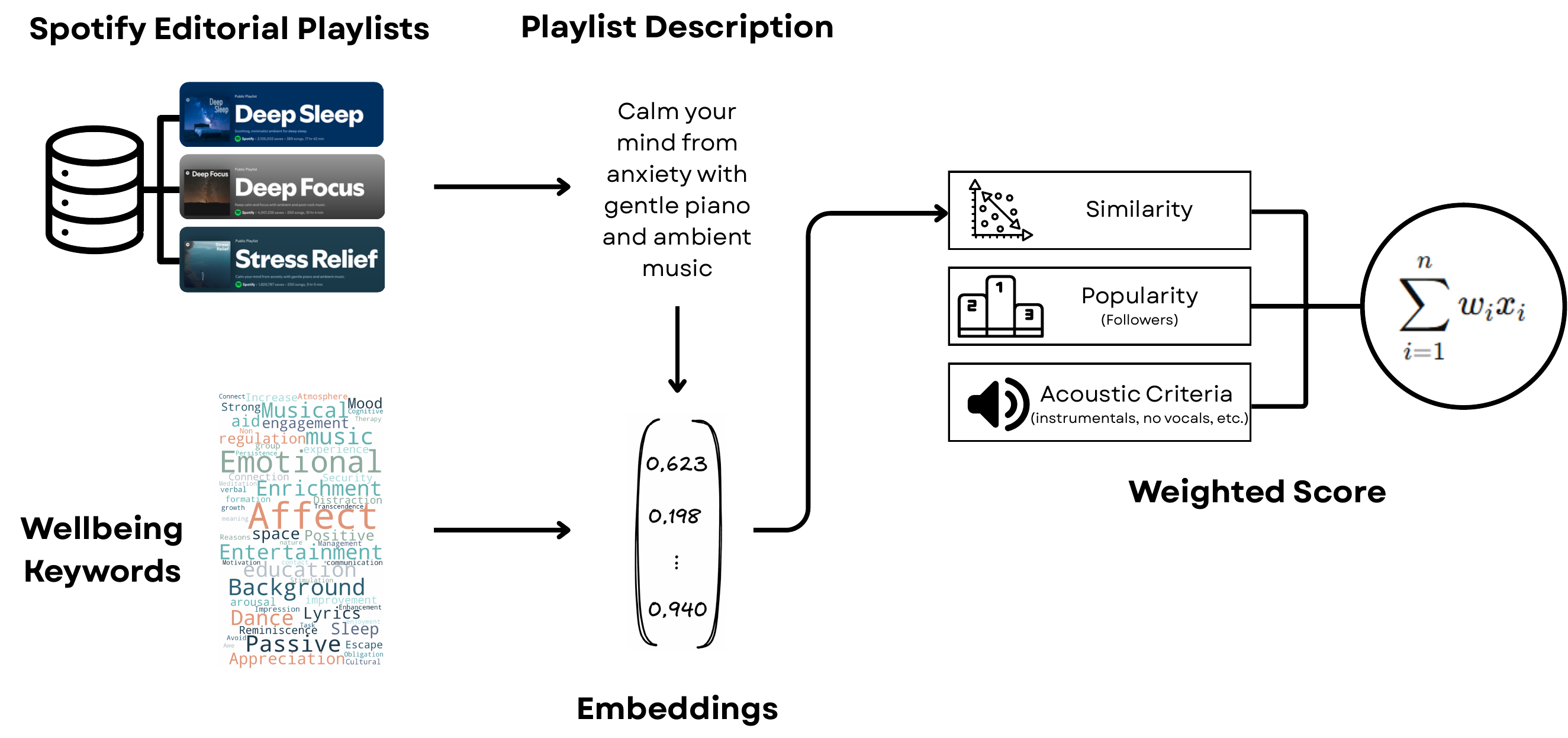}
  \caption{Schematic overview of the stimuli selection process. Text embeddings are extracted from playlist descriptions and wellbeing keywords to compute similarity measures, which are combined with additional criteria to derive a final weighted score, used to select the best-ranked playlist per function. }
  \label{fig:stimuli_selection}
\end{figure}

To support the temporal alignment of physiological data with music listening activity and their association with participant feedback, we designed and developed a custom monitoring system. The infrastructure also incorporated a web-based platform for participant enrolment, resource allocation, administration of validated questionnaires, and tracking of playlist engagement. The system logs participant playback state through periodic API calls to record the onset and offset of listening episodes and schedules short surveys to assess perceived music effects. These records were subsequently linked to each participant’s physiological data and further associated with individual difference and well-being measures.

At enrolment, participants attended a laboratory onboarding session during which written informed consent was obtained and a unique participant identifier was generated. The wearable device, featuring PPG, EDA, skin temperature, and triaxial acceleration sensors, was pre-sanitised, configured, worn on the non-dominant hand, and paired with the mobile application, which periodically uploads physiological data to the cloud storage. Participants received instructions on device use, highlighting continuous wearing time and to recharge it while showering or when the battery is low. They were subsequently registered on the study platform to complete baseline questionnaires and assigned a dedicated Spotify account preloaded with the curated playlists. Finally, the experimenter ensured proper device pairing, application configuration, music listening tracking and playlist access prior to initiating the 21-day monitoring period.

\subsection{Procedure} Figure 1 illustrates the overall data collection procedure. After enrolment, participants undertook a 21-day in-the-wild monitoring period during which they maintained their usual daily activities. The wearable device was worn continuously on the non-dominant hand to capture multimodal physiological signals throughout the study. A study-specific Spotify account containing 12 curated playlists was provided, and participants were instructed to treat it as their personal account throughout the study period. However, only listening activity associated with the study playlists are included in the dataset.

Participants were encouraged to engage with the playlists whenever they wished to benefit from their intended therapeutic effects. No prescriptive instructions regarding specific activities were provided, beyond remaining in the relevant context (e.g., office work, relaxation, meditation) while listening to the study playlists. They were asked to listen to each 30-minute playlist at least once from start to finish with minimal interruptions, ideally using headphones when others were present or speakers when alone.

The monitoring system automatically logged playback activity, including track start and end times, as well as associated metadata. Detected listening activity prompted a short email-based survey delivered at specific time of the day (09:00 or 21:00), with the purpose of reducing notification fatigue and ensuring high response rate. The survey consisted of a single item assessing whether the playlist supported its intended function (e.g., reduce distraction, relaxation, mental clarity, or overall helpfulness). Upon completion of the 21-day period, participants received an automated email confirming the end of the study and were invited to return the device and complete post-study questionnaires. After returning the equipment and finalising assessments, participants were thanked and compensated.

\subsection{Collected Data}
The corpus comprises three main components: (1) \textbf{Participants' physiological data} including continuous raw signals and aggregated biomarkers per minute provided by the device manufacturer; (2) \textbf{Psychometric data} including demographics, participant characteristics (personality, music preferences, and musical sophistication), and measures of affect, perceived stress, and depressive symptoms; and (3) \textbf{Music stimuli} and related data covering music listening activity and computed acoustic features. Below we describe each component in detail.
%\TG{Are we not going to talk about each component? That would make more sense.} \BGS{I followed the structure of Fahim's previous Scientific Data submissions and the journal guidelines. The collected data is a subsection within the Methods section describing how each data component was collected, whereas Data Records explains the files, formats, variables, and repository structure.}

\subsubsection*{Participants' physiological data} 
%subsubsection numbering does not look good, removing.
Physiological data were collected using the Empatica EmbracePlus (\href{https://www.empatica.com/en-gb/embraceplus/}
{\nolinkurl{https://www.empatica.com/en-gb/embraceplus/}}), a wrist-worn sensor platform designed for continuous ambulatory monitoring. The device records multimodal physiological signals, including PPG, EDA, skin temperature, and triaxial acceleration, enabling long-term data acquisition without substantially disrupting participants’ daily activities. Sensor specifications are summarised in Table 2. 
\begin{table}[t]
\centering
\footnotesize
\setlength{\tabcolsep}{3.5pt}
\renewcommand{\arraystretch}{1}
\label{tab:sensor_specifications}
\caption{Sensor specifications of the Empatica EmbracePlus device used for physiological data acquisition.}
\vspace{2mm}
\begin{tabular}{p{2.0cm} p{2.5cm} p{2.0cm} p{7.6cm}}
\toprule
\textbf{Sensor} & \textbf{Measurement Range} & \textbf{Sampling Rate} & \textbf{Description} \\
\midrule

PPG 
& -- 
& 64 Hz
& Photoplethysmography signal used to estimate blood volume pulse and derive cardiovascular-related biomarkers. \\

EDA 
& 0.01--100 $\mu$S 
& 4 Hz 
& Electrodermal activity measuring skin conductance changes associated with sympathetic nervous system activation. \\

IMU
& $\pm$16g, $\pm$2000dps 
& 64 Hz 
& Triaxial acceleration capturing body movement and physical activity levels. \\

Temperature 
& 0--50$^\circ$C 
& 1 Hz 
& Peripheral skin temperature measured at the wrist. \\

\bottomrule
\end{tabular}
\end{table}
The EmbracePlus operates within the Empatica Health Monitoring ecosystem, comprising the CareLab (\href{https://support.empatica.com/hc/en-us/articles/9843547409821-Download-the-Care-App}
{\nolinkurl{https://support.empatica.com/hc/en-us/articles/9843547409821-Download-the-Care-App}}) mobile application, Empatica Cloud storage, and the web-based Care Lab Portal. The device connects to participants’ smartphones via the CareLab app, which uploads physiological recordings periodically to Empatica Cloud. Participants were required to install the CareLab app and keep it running in the background to ensure continuous data transfer. Study-specific login credentials were generated through the Care Lab Portal, which also enables participant management and monitoring of derived biomarkers.

Raw physiological signals are stored in Avro format and can be accessed for further preprocessing and analysis. In addition, minute-level digital biomarkers computed using Empatica’s proprietary algorithms, are also included in the dataset.

\subsubsection*{Psychometric data}
Evidence from previous research indicates that emotional responses to music may be influenced by personality \cite{juslin2008experience}, musical training \cite{lehmann1997research}, and music preference \cite{north1997liking}. To account for these factors, validated questionnaires were administered at registration to capture relevant individual traits, alongside demographic information. Additionally, to evaluate potential changes associated with sustained music engagement, overall well-being was measured at both registration and study completion. Table 3 summarises the questionnaires administered throughout the study, including enrolment measures of demographic characteristics and individual differences, as well as pre/post assessments of mood and emotional state, perceived stress, and depressive symptom severity.

\begin{table}[!b]
\centering
\footnotesize
\setlength{\tabcolsep}{4pt}
\renewcommand{\arraystretch}{1.35}
\caption{Summary of questionnaires used to assess participant demographics, individual traits, and pre/post psychological measures.}
\label{tab:questionnaires}
\vspace{2mm}
\begin{tabular}{
>{\raggedright\arraybackslash}m{2.2cm}
>{\raggedright\arraybackslash}m{3.4cm}
>{\raggedright\arraybackslash}m{3.8cm}
>{\centering\arraybackslash}m{0.8cm}
>{\raggedright\arraybackslash}m{3.0cm}
>{\raggedright\arraybackslash}m{3.6cm}
}
\toprule
\textbf{Type} &
\textbf{Full Name} &
\textbf{Information Collected} &
\textbf{Items} &
\textbf{Response Range} &
\textbf{Overall Score} \\
\midrule

Demographics &
N/A &
Personal details, work context, music listening habits &
22 &
Varied formats &
N/A \\

Personality &
Ten Item Personality Inventory (TIPI) \cite{gosling2003very} &
Big Five personality traits &
10 &
1--7 (Disagree to Agree) &
Average score for each trait \\

Music Sophistication &
Goldsmiths Musical Sophistication Index (Gold-MSI) \cite{mullensiefen2014musicality} &
Musical skills, experience, and engagement &
38 &
1--7 (Disagree to Agree) &
Five subscales and composite index \\

Music Preference &
Short Test of Music Preference (STOMP) \cite{rentfrow2003re} &
Genre preference ratings &
23 &
1--7 (Dislike to Like) &
Four broad preference dimensions \\

Affect &
Positive and Negative Affect Schedule (PANAS) \cite{watson1988development} &
Positive and negative emotional states &
20 &
1--5 (Very Slightly to Extremely) &
Sum of responses for each subscale \\

Perceived Stress &
Perceived Stress Scale (PSS) \cite{cohen1983global} &
Stress experienced over the past month &
14 &
0--4 (Never to Very Often) &
Total score; higher values indicate greater stress \\

Depression Severity &
Patient Health Questionnaire (PHQ-9) \cite{kroenke2001phq} &
Depressive symptoms over the past two weeks &
9 &
0--3 (Not at All to Nearly Every Day) &
Sum of item scores indicating depression severity \\

\bottomrule
\end{tabular}
\end{table}

\subsubsection*{Music stimuli}
Music listening activity was continuously logged through automated Spotify API (\href{https://developer.spotify.com/documentation/web-api}
{\nolinkurl{https://developer.spotify.com/documentation/web-api}}) calls, capturing playback timestamps, track metadata, playback duration, context, device type, and whether tracks were played in full. Playback records were retrieved every 30 seconds and cross-validated against Spotify timestamps to ensure temporal accuracy. 

Subjective responses were collected through an automated email-based survey system. If participants engaged with study playlists within the preceding 12 hours, they received a single-item confirmatory question assessing whether the playlist supported its intended function. Responses were recorded as binary (True/False) indicators and linked to the corresponding listening session. 

Due to copyright restrictions, raw audio files of the stimuli cannot be distributed. Instead, our dataset provides Spotify URLs and tracked metadata to enable stimulus identification, acknowledging that availability may change over time. To support reproducible analysis, both static and dynamic acoustic features are included. Low-level descriptors were computed using short-time Fourier transform (STFT) following a similar methodological approach as in Hu et al. \cite{hu2022detecting}. Acoustic features were extracted using a standardised Python-based audio analysis pipeline using the \textit{Librosa} (\href{https://librosa.org/doc/latest/index.html}
{\nolinkurl{https://librosa.org/doc/latest/index.html}}) package with default parameters (i.e., \textit{n\_fft = 2048, hop\_length = 512, win\_length = 2048}, etc.).  In addition, complexity was quantified using audio compressibility ratio as introduced in \cite{kumar2014affective}. Global (static) features were obtained by aggregating frame-level descriptors across the full track duration computing mean and standard deviation as statistical functionals.

To characterise temporal evolution, dynamic features were computed by segmenting each audio signal into non-overlapping 0.25 second windows and using mean aggregation. Non-overlapping segmentation was applied to avoid temporal leakage or filtering effects between adjacent windows.

\section{Data Records}

All participant data were anonymised using alphanumeric identifiers following the format P\#\# (e.g., P01). The dataset is structured into three primary directories: \texttt{physiology/}, \texttt{psychometrics/}, and \texttt{stimuli/}. Further details on each data modality are provided below.

\subsection{Physiology Data}

Physiological data are available in the \texttt{physiology/} directory and include raw sensor recordings, minimally processed signals, and one-minute aggregated digital biomarkers collected using the EmbracePlus device. Raw recordings correspond to the original serialised files downloaded from Empatica Cloud and are stored in \texttt{.avro} format. Processed physiological signals are provided in \texttt{.parquet} format and include blood volume pulse, electrodermal activity, skin temperature, triaxial acceleration, and systolic peak timestamps. These files were aligned to a 00:00 reference timeline and corrected for cross-day discontinuities. In addition, periods with wearing-detection percentages below 75\% were filtered out to reduce the inclusion of artefacts and invalid measurements collected while the device was not worn.

One-minute aggregated digital biomarkers are provided in \texttt{.csv} format and correspond to pre-processed physiological and behavioural measures derived from Empatica’s proprietary and open-source algorithms from the raw sensor data. These biomarkers include measures related to cardiovascular activity, pulse rate variability, respiratory dynamics, electrodermal activity, skin temperature, physical activity, sleep, gait, and wearing detection, among others. Both processed physiological signals and aggregated biomarker files consist of timestamped measurements paired with their associated physiological values. Altogether, the BEATS dataset includes data from 30 participants monitored over 21 days, yielding 10,725 hours of raw physiological recordings from PPG, EDA, skin temperature, and triaxial accelerometer. Figure 3 illustrates an example music listening session aligned with multimodal physiological signals.
\begin{figure}[ht]
  \centering
\includegraphics[width=\linewidth]{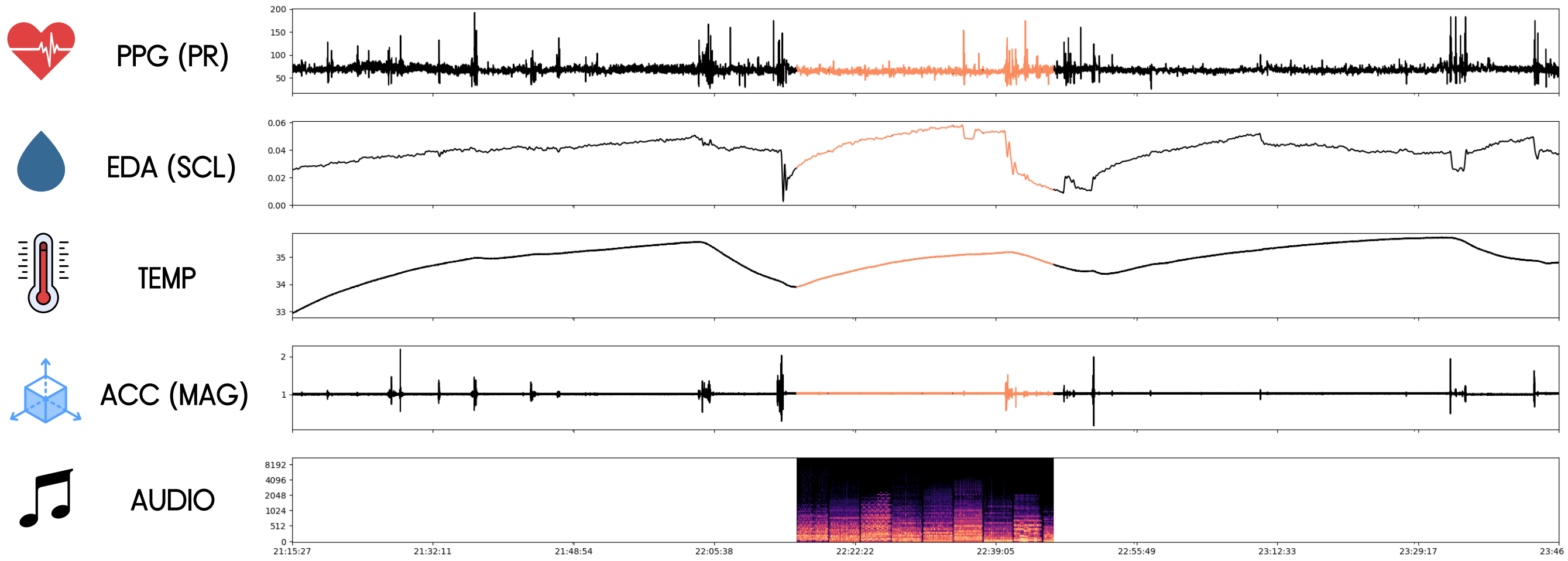}
  \caption{Sample snippet from our datasets showing  multimodal physiological signals aligned with music listening activity. Physiological signals are recorded during music playback is highlighted in orange, while pre- and post-listening periods are shown in black.} 
  \label{fig:physio-aligned-music}
\end{figure}
\subsection{Demographics and Psychometrics} Self-reported participant information and questionnaire scores are provided in the \texttt{psychometrics/} directory as separate files for each questionnaire. More specifically, demographic information includes participant characteristics such as non-dominant hand (device placement), recruitment period, work context, and music listening habits. Furthermore, individual-difference measures include trait-level assessments of personality (TIPI), music preferences (STOMPR), and general music sophistication (Gold-MSI), with scores reported for each questionnaire dimension. Additionally, well-being measures include pre- and post-study assessments of perceived stress (PSS), depressive symptoms (PHQ-9), and affect (PANAS). Figure 4 summarises the personality, music preference, musical sophistication, and wellbeing questionnaire scores included in the dataset.

\begin{figure}[ht]
  \centering
\includegraphics[width=\linewidth]{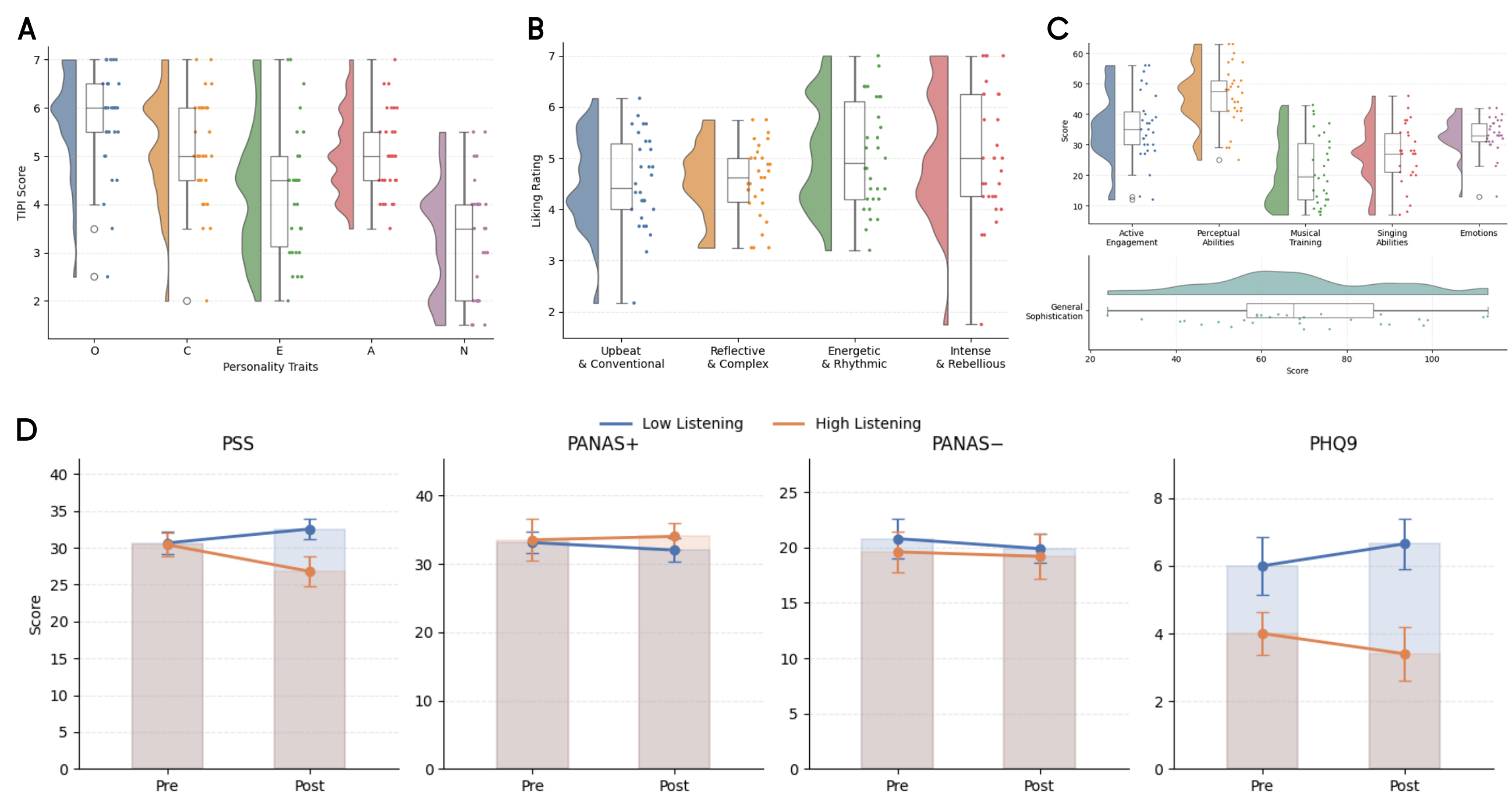}
  \caption{Overview of self-reported psychometric measures included in our dataset. (A) Distribution of TIPI personality trait scores of participants. (B) Distribution of STOMPR music preference scores across music preference dimensions. (C) Distribution of Gold-MSI scores, including individual dimensions and general sophistication. (D) Pre- and post-study psychological measures grouped by high and low music listening activity, including perceived stress, positive/negative affect, and depression severity.}
  \label{fig:psychometrics-scores}
\end{figure}

\vspace{-1.0em}
\subsection{Music Listening and Stimuli Data} Music-related data are provided in the \texttt{stimuli/} directory and these include stimuli metadata and pre-computed acoustic features. Static features are provided as 234 descriptors computed using mean and standard deviation aggregation functionals. Dynamic acoustic features comprise 116 low-level descriptors extracted at 0.25 second resolution (4 Hz) and aggregated using mean, enabling time-aligned analyses with physiological signals.

The dataset also includes detailed music listening records capturing real-world listening behaviour, with contextual information and precise Unix timestamps (microseconds) marking the start and end of each listening sample, as well as self-reports collected after listening sessions. To facilitate data integration, common field names are used consistently across stimuli metadata, music listening records, and self-reports.

Figure 5 summarises the music listening activity and post-listening self-reports associated with the study stimuli. In total, 262.2 hours of music listening were aligned with physiological recordings across 844 listening sessions and 3 listening contexts involving 117 curated tracks. Goal-Attainment accounted for the largest proportion of listening time (176.4 h), followed by Affective (54.4 h) and Eudaimonic (31.8 h) contexts. Post-listening surveys achieved a response rate of approximately 90.8\%, with most responses indicating that the music supported the intended listening purpose.

\begin{figure}[ht]
  \centering
\includegraphics[width=\linewidth]{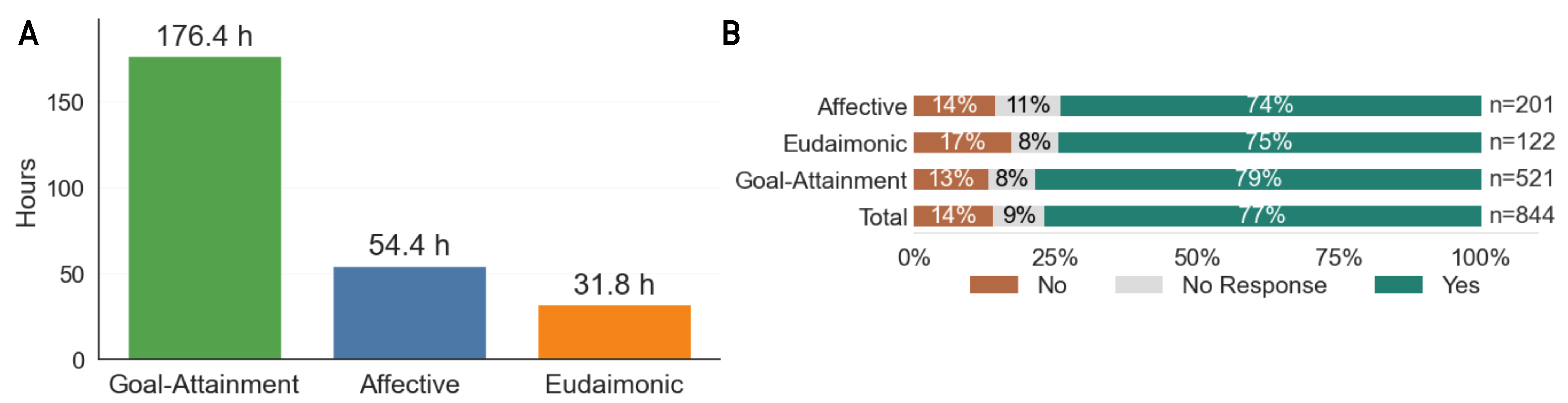}
  \caption{Listening engagement and subjective evaluations of the study stimuli. (A) Total listening time (hours) across music functions recorded during sessions overlapping with physiological measurements. (B) Participant responses to post-session surveys evaluating whether the music supported the intended listening purpose.}
  \label{fig:listening-self-reports}
\end{figure}

\vspace{-1.5em}
\section{Technical Validation}

In this section, we present analyses designed to evaluate the technical validity of the dataset with respect to the musical stimuli and physiological signals.

\subsection{Acoustic Profiles of Stimuli}
To validate the musical stimuli, we examined whether each music function in the dataset exhibits a distinct acoustic profile. A supervised learning approach was adopted to identify salient discriminative features across the three music functions using low-level acoustic descriptors extracted from the audio stimuli. Given the relatively small sample size, we trained a logistic regression classifier with L1 regularisation using three-fold cross-validation, achieving a mean F1 score of 0.785. The L1 penalty drove coefficients associated with less informative features towards zero, highlighting the most influential descriptors while reducing dimensionality to 10 features, primarily related to rhythmic and timbral characteristics.

\begin{figure}[ht]
  \centering
\includegraphics[width=\linewidth]{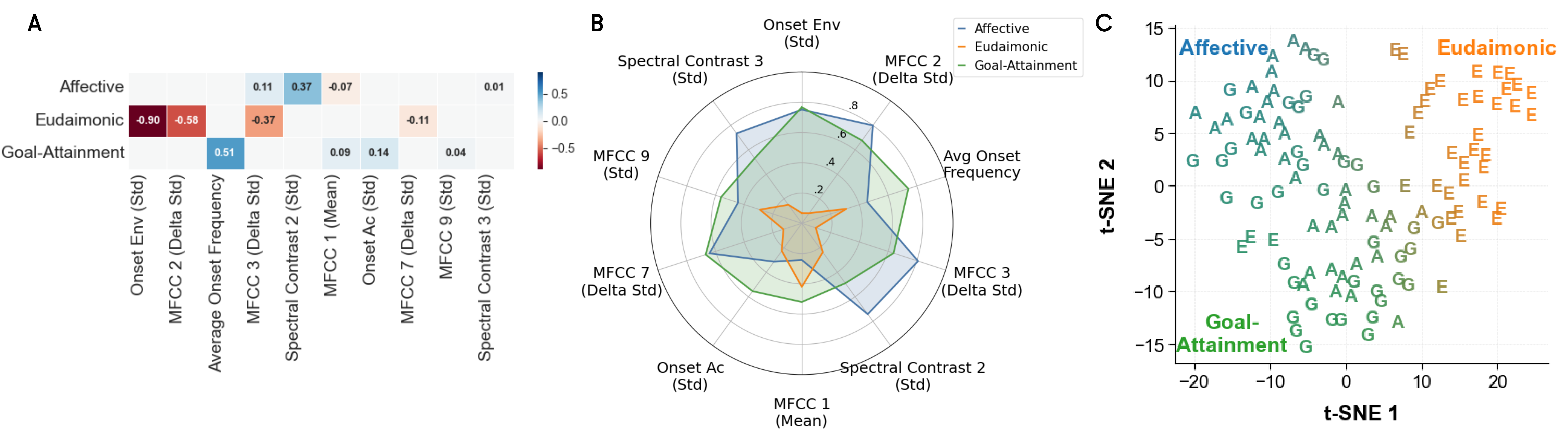}
  \caption{Acoustic characterisation of the musical stimuli. (A) Logistic regression feature weights identifying discriminative acoustic descriptors across music functions. (B) Radial plot comparing the selected descriptors for Affective, Eudaimonic, and Goal-Attainment music, illustrating distinct acoustic profiles. (C) Two-dimensional t-SNE embedding of the selected features showing the acoustic distribution of Affective, Eudaimonic, and Goal-Attainment music.}
  \label{fig:stimuli-validation}
\end{figure}

Figure 6 summarises the resulting feature weights and contrasts associated features across music functions using a radial plot, suggesting distinct acoustic profiles for each category. Notably, the descriptors with the largest weights correspond to spectral contrast for Affective music, onset strength envelope for Eudaimonic music, and average onset frequency for Goal-Attainment music. Spectral contrast quantifies the difference between spectral peaks and valleys across frequency bands, while the onset strength envelope reflects the rate of change in the power spectrum between consecutive frames, and the average onset frequency approximates the number of note onsets per second. The weights assigned to these descriptors suggest that Goal-Attainment stimuli in the dataset contain a higher density of note events, Affective music exhibit greater variation in timbral texture, and Eudaimonic tracks are characterised by smoother and less abrupt musical progression over time.

To further inspect the acoustic structure of the selected descriptors, we visualised the ten-feature representation using a two-dimensional embedding obtained with the t-SNE algorithm. Class probabilities derived from the trained logistic regression model were mapped onto the embedding. The resulting representation reveals continuous gradients between categories, revealing that Goal-Attainment songs are acoustically closer to Affective than to Eudaimonic music.

\subsection{Physiological Variation across Listening Periods}

We next examined physiological markers during music listening by contrasting their variation across pre-, during-, and post-listening periods. Participants were free to engage with the study playlists at times that best suited their needs, while wearing the wristband as consistently as possible. This allowed the continuous acquisition of physiological data both during and outside music listening episodes. For each participant, we identified all music listening episodes and corresponding one-hour windows preceding and following each episode. Music listening episodes were defined as continuous intervals during which study stimuli were played, with durations varying across sessions. To enable comparisons across periods despite variable episode lengths, all time segments were partitioned into non-overlapping one-minute windows. From each short-term window, physiological features were extracted from heart rate variability (HRV), electrodermal activity (EDA), skin temperature, and triaxial accelerometry.

To account for inter-individual variability, we first established a daily physiological reference for each participant, defined as the most stable physiological state observed within a given day. Baseline windows were identified using a data-driven approach combining two criteria: (i) the participant was awake and wearing the device correctly, and (ii) modality-specific conditions reflecting low and stable physiological activity. For HRV, candidate segments were selected using a 15-minute rolling mean to identify low heart rate windows overlapping with low motion. For EDA, segments were defined as periods of low tonic activity overlapping with the same low-motion constraint and stable skin temperature. Low physiological activity was determined as values below the 30th percentile of the daily distribution, while stable temperature to samples within the daily interquartile range. Afterwards, the daily baseline mean and standard deviation were computed for each participant and used to standardise all window-level features within the same day to bring all participant samples onto a common scale for comparison across periods.

\begin{figure}[ht]
  \centering
\includegraphics[width=\linewidth]{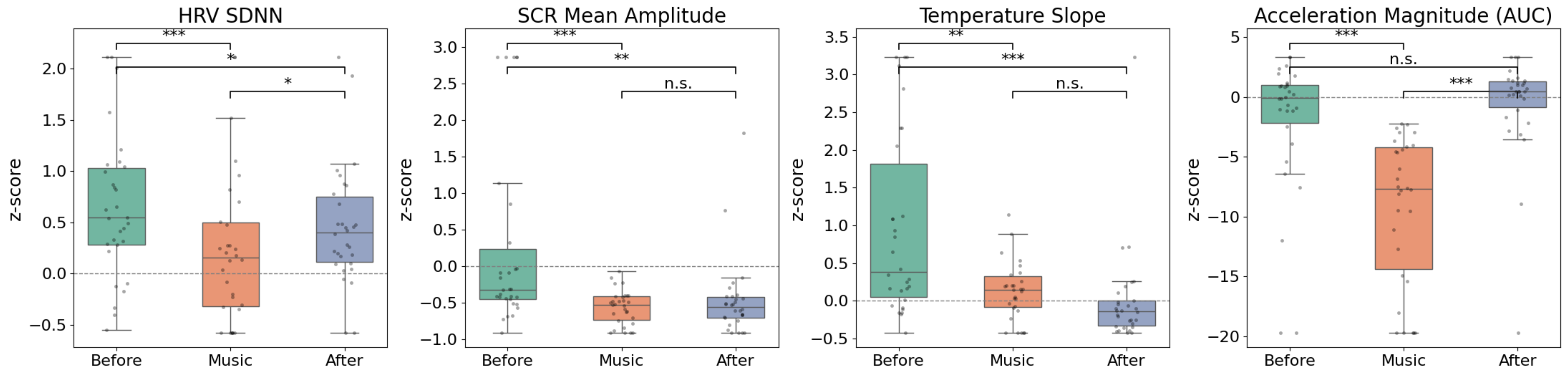}
  \caption{Distributions of normalised physiological markers across the three temporal segments. Values are aggregated at the participant level following day-wise standardisation. Box plots illustrate the distribution of each marker, with statistical significance of pairwise comparisons indicated (*p$<$0.05, **p$<$0.01, ***p$<$0.001; n.s., not significant).}
  \label{fig:music-vs-nomusic}
\end{figure}

Window-level features were subsequently mean-aggregated within each participant and period (before, during, after), yielding a single representative value per participant for each condition. Paired t-tests were conducted on participant-level samples (n = 30) to assess differences across periods for each physiological marker. Figure 7 shows the distribution of selected markers across the three temporal segments. The resulting distributions indicate that HRV, EDA, and temperature shift closer to or below individual baseline levels during music listening, while motion is significantly lower during listening relative to before and after periods. Interestingly, electrodermal activity and temperature remain closer to baseline after listening, suggesting sustained physiological regulation beyond the listening session. However, these observations should be interpreted within the context of real-world data collection, where several confounding factors were not controlled. Additionally, aggregation at the participant level may obscure important fine-grained temporal dynamics. Despite these limitations, the observed trends motivate further investigation.

\vspace{-1.0em}
\section{Usage Notes}

We recommend working with the minimally processed physiological data provided in the dataset, although users are encouraged to develop and apply their own preprocessing pipelines and analytical approaches. For custom physiological signal processing, NeuroKit (\href{https://neuropsychology.github.io/NeuroKit/}{\nolinkurl{https://neuropsychology.github.io/NeuroKit/}}) and cvxEDA (\href{https://github.com/lciti/cvxEDA}{\nolinkurl{https://github.com/lciti/cvxEDA}}) may be useful resources for deriving physiological measures and decomposing electrodermal activity signals. Users interested in working directly with the original Empatica recordings may follow Empatica's official documentation for accessing AVRO files using Python (\href{https://support.empatica.com/hc/en-us/articles/17405877853981-How-to-access-Avro-files-with-Python}{\nolinkurl{https://support.empatica.com/hc/en-us/articles/17405877853981-How-to-access-Avro-files-with-Python}}).

For dataset inspection, quality control, participant selection, or session filtering, the one-minute aggregated physiological biomarkers can provide useful summaries of participant state throughout the day, including wearing detection, sleep periods, and intervals of low physical activity. Given the in-the-wild nature of the data collection, physiological responses and listening behaviours may be affected by day-to-day activities, environmental conditions, and individual circumstances that were not experimentally controlled. Users should consider these sources of variability when analysing the dataset and interpreting findings.

\section{Data and Code Availability}

The BEATS dataset is available through the project website at \href{https://beats-dataset.github.io/}{\nolinkurl{https://beats-dataset.github.io/}}\footnote{The dataset repository link will become available upon completion of the peer-review process.}. It contains anonymised physiological, psychometric, and music-related data, including listening activity, self-reports, stimuli metadata, and acoustic features, as described in the Data Records section.

%\section*{Code Availability}

The source code associated with the BEATS dataset is publicly available at \href{https://github.com/gtzben/beats-dataset}{\nolinkurl{https://github.com/gtzben/beats-dataset}}.

\section*{Author Contributions}

B.G.S., T.G., and F.K. contributed to the design of the study protocol. B.G.S. developed the software, conducted the experiments, collected the data, and prepared the first draft of the manuscript. T.G. and F.K. supervised the project. All authors contributed to manuscript revision and approved the final version of the manuscript.

\section*{Competing Interests}
The authors declare no competing interests.

\section*{Acknowledgements}
This work was supported by the UKRI Centre for Doctoral
Training in Socially Intelligent Artificial Agents, Grant Number
EP/S02266X/1. 

\end{document}